\documentstyle[prb,aps,multicol]{revtex} 
\tighten
\begin{document}
\draft
\title{Reentrant Condensation of DNA induced by Multivalent Counterions}
\author{T. T. Nguyen, I. Rouzina* and B. I. Shklovskii}
\address{Theoretical
Physics Institute, University of Minnesota, 116 Church St. Southeast,
Minneapolis, Minnesota 55455  and
*Department of Biochemistry, University of Minnesota, 1479 Gortner Ave. St. Paul,
Minnesota 55108}
\maketitle

\begin{abstract}
A theory of condensation and resolubilization of a dilute DNA 
solution with growing concentration of multivalent cations, $N$
is suggested.
It is based on a new theory of screening of a macroion by multivalent
cations,
which shows that due to strong cation correlations at the surface of DNA 
the net charge of DNA changes sign at some small concentration of
cations $N_0$.
DNA condensation takes place in the vicinity of $N_0$, 
where absolute value of the DNA net charge is small and the correlation 
induced short range attraction dominates the Coulomb repulsion.
At $N > N_0$ positive DNA should move in the oppisite
direction in an electrophoresis experiment. 
From comparison of our theory with experimental values of 
condensation and resolubilization thresholds
for DNA solution containing Spe$^{4+}$,
we obtain that $N_0$ = 3.2~mM and that the
energy of DNA condensation per nucleotide is 
$0.07~k_B T$. 
\end{abstract}

 \begin{multicols}{2}

\section {Introduction}
In the last several years there has been a revival of interest in the
phenomenon of DNA condensation with multivalent cations. The reason for
this
interest is the general effort of the scientific community to develop
effective ways of gene delivery
 for the rapidly growing field of genetic therapy. The DNA compaction
should
be fast, effective, easily
 reversible and should not damage the DNA double helix. All of these
conditions are fulfilled in DNA
condensation with multivalent cations, such as CoHex$^{3+}$, naturally
occurring polyamines Spd$^{3+}$,
 Spe$^{4+}$ and their analogs which
are known to bind to DNA in the predominantly nonspecific electrostatic
manner. The DNA condensates
 obtained this way are indeed closely packed arrays of parallel DNA
strands.
It was shown that the helical structure of the B-DNA is not perturbed
within such condensate, and
 that the reaction is easily reversed by the addition of monovalent
salt,
or simply dilution of the
 solution with water. Cations with larger and more compact charge are
more effective in condensing
the DNA. This also suggests an electrostatic mechanism of the DNA
condensation
with multivalent cations.
 During about twenty years of research, a significant amount of
information
on DNA condensation has been accumulated.
 For long DNA, as the concentration of $Z$-valent cations grows
condensation happens abruptly at some critical concentration, $N_c$
which
depends on the charge of cations and the concentration of monovalent
salt
$n$. A comprehensive review of the
experimental and theoretical results for $N_c$
can be found in Refs.~\onlinecite{Vic1,Vic2}.

The intensive study of the last few years revealed completely new features
of DNA condensation with multivalent
 cations~\cite{Thomas,Livolan96,Pelta,Raspaud}.
It was discovered that when the concentration of cations grows far beyond
$N_c$
to some new
critical value $N_{d} \gg N_c$, DNA dissolves and returns to the
solution.
This reentrance condensation behaviour is schematically shown on Fig.~1.
For a long DNA and small $n$ both
transitions are very sharp and the ratio of $N_{d}/N_c$ can be as large
as
$10^4$. Remarkably, the decondensation threshold $N_{d}$ is almost
totally
independent on
the monovalent salt concentration, $n$. On the other hand, the
condensation threshold, $N_c$, grows with
increasing $n$.
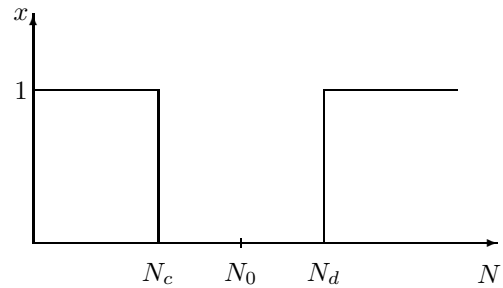
\begin{figure}[h]
\setlength{\unitlength}{0.140900pt}
\ifx\plotpoint\undefined\newsavebox{\plotpoint}\fi
\sbox{\plotpoint}{\rule[-0.200pt]{0.400pt}{0.400pt}}%
\begin{picture}(1500,900)(0,0)
\font\gnuplot=cmr10 at 10pt
\gnuplot
\sbox{\plotpoint}{\rule[-0.200pt]{0.400pt}{0.400pt}}%
\put(438,163){\makebox(0,0){$N_c$}}
\put(883,163){\makebox(0,0){$N_{d}$}}
\put(661,163){\makebox(0,0){$N_0$}}
\put(1328,163){\makebox(0,0){$N$}}
\put(71,860){\makebox(0,0){$x$}}
\put(71,655){\makebox(0,0){$1$}}
\put(104,245){\vector(0,1){615}}
\put(104,245){\vector(1,0){1246}}
\put(661,233){\line(0,1){24}}
\put(104,655){\usebox{\plotpoint}}
\put(104.0,655.0){\rule[-0.200pt]{47.500pt}{0.400pt}}
\put(438.0,245.0){\rule[-0.200pt]{0.400pt}{58.000pt}}
\put(438.0,245.0){\rule[-0.200pt]{107.200pt}{0.400pt}}
\put(883.0,245.0){\rule[-0.200pt]{0.400pt}{58.000pt}}
\put(883.0,655.0){\rule[-0.200pt]{50.461pt}{0.400pt}}
\end{picture}
\caption{Schematical illustration of the reentrant condensation. The
fraction $x$ of DNA
molecules in solution is plotted as a function of logarithm of the
cation
concentration}
\end{figure}

It has been understood for some time that, due to correlations between
multivalent cations at the DNA
surface~\cite{Oosawa,Gulbrand,Roland,Rouz96,Bruinsma,And,Parsegian,Levin,Shklov98},
 two DNA molecules experience
a short range attraction, which can lead to condensation. Monovalent
ions
are much less correlated and do not provide
any attraction. Conventional explanation
for the condensation threshold $N_c$ is that it is just 
the bulk concentration of $Z$-valent cations, $N = N_r$,
at which they replace monovalent ones and begin to produce a short range
attraction.
However, from this point of view it is very difficult to understand
why at large $N = N_{d}$, DNA molecules go back to the solution.

The above explanation also does not take into account the net charge of
DNA.
The nonlinear Poisson-Boltzmann equation predicts~\cite{Manning,Zimm} a
value for the net linear charge density of DNA, $\eta^*$,
which includes the bare charge of DNA
and the charge of cations bound at the very
surface of DNA with energy larger than $k_BT$. This charge
is negative and does not depend on $N$.
At  $N > N_r$, two situations are possible. The energy of the
Coulomb
repulsion
can be smaller than the energy of the short range attraction. Then DNA condenses
at
$N = N_r$
but never dissolves back. If the energy of the Coulomb repulsion is larger
than the energy of short range attraction, condensation does not happen at
all.
Both possibilities contradict experiments
and therefore reentrant condensation can not be explained in the
Poisson-Boltzmann aproximation.

In this paper we propose an explanation for the  reentrant condensation
based on a new
theory of screening of macroions by multivalent cations, which
emphasizes the
strong correlations of multivalent
cations at the surface of DNA.
We explain why the condensed phase exists only within the limited range,
$N_{c} \leq N \leq N_{d}$ and calculate both $N_{c}$ and
$N_{d}$.

It was shown in Ref.~\onlinecite{Perel,Shklov99}, that the strong repulsion
between multivalent cations leads to their strong lateral correlations.
The resulting strongly correlated liquid of multivalent cations at the
surface of DNA
has a large negative
chemical potential, which describes the additional purely
electrostatic binding of cations to the surface. This,
in turn, leads to an exponentially small
concentration of cations, $ N_{0}$, above
the surface. ($N_0 > N_r$, if $ n$ is not unrealistically large).
According to Ref.~\onlinecite{Perel,Shklov99}, when the bulk
concentrations
of multivalent cations, $N$, grows above
$N_r$, the net negative charge $\eta^*$ decreases in absolute value and  crosses
zero at $N = N_{0}$. At  $N > N_{0}$,
the net charge  becomes $positive$ and continues to grow with
$N$.
This effect is called charge inversion. It is worth noting here, that
the
charge
inversion is not a result of some specific chemosorbtion~\cite{Leikin},
but
rather a direct consequence
of purely electrostatic interactions of multivalent cations.

At  $N = N_{0}$ there is no Coulomb repulsion at all, so that
the short range attraction dominates and leads to condensation.
This is actually the optimal situation for condensation.
It is obvious that there is a range of $N$
around $N_{0}$ where attraction still dominates and DNA condences.
 Condensation thresholds $N_c$ and $N_d$ then are determined by the
condition that
the energy of the short range attraction is equal to the energy of
Coulomb repulsion of negative and positive DNA molecules respectively.
This means that the concentration $N_{0}$ is located inside the window
$N_{c}
\leq N \leq N_{d}$ (see Fig. 1)
and the width of the window grows with the strength of the short range
attraction.

Below we present an analytical theory, which calculates both critical
concentrations
 $N_{c}$ and $N_{d}$ in terms of the two main
physical parameters of the system: boundary concentration of multivalent
cations $N_{0}$
and the energy, $\varepsilon$, of
the DNA binding within the DNA bundle
per nucleotide.

At a small concentration of monovalent salt, $n$, the condensation
threshold $N_c$ calculated in our theory
is larger than the replacement concentration $N_r$. This means that
the replacement of monovalent cations happens
before the condensation and by itself can not lead to the condensation
while Coulomb forces are still strong.
On the other hand, if $n$ is so large that $N_r $ is larger than the
value of
$N_c$ calculated in our theory (which assumes that replacement has
happened), the condensation
actually happens only at $N = N_{r}$. Thus, a large
concentration
of monovalent ions, $n$, acts as
a "curtain". It eliminates a part of the window predicted by our theory.
We
show below that in
the experiment of Ref.~\onlinecite{Thomas}
the "curtain effect" starts to work only at $n > 50$~mM.

Two central parameters of our theory, $N_{0}$ and $\varepsilon$,
can be calculated for a simple model of DNA as an uniformly charged
cylinder~\cite{Shklov98,Perel,Shklov99}.
They can be also directly measured in independent experiments. The
energy
$\varepsilon$ was measured in Ref.~\onlinecite{Rau92}.

On the other hand, one can obtain these parameters from the experimental
values of
$N_{c}$ and $N_{d}$. Below we use condensation and resolubilization for
DNA
with Spe$^{4+}$
to estimate $\varepsilon$ and $N_{0}$. We obtain $\varepsilon =
0.07~k_BT$
and $N_{0} = 3.2~mM$.
The first number reasonably agrees with experimental data\cite{Rau92}.

Thus far we have talked about long DNA.
For short DNA fragments, one should take into account the mixing
entropy of DNA molecules which makes
their solution more stable and makes the window between $N_{c}$ and
$N_{d}$
smaller.
We calculate the phase diagram for the condensation of DNA olygomers
consisting
of $L/b$ bases ($L$ is
the length of olygomer, $b = 1.7~\AA$ is length of double helix per
phosphate). This phase diagram
is shown in Fig.~2 on the plane $(N, y)$, where
\begin{equation}
y = {{k_BT b}\over{\varepsilon L}}\ln {C_{max}\over {C}}
~~~, \label{y}
\end{equation}
 $C$ is the concentration of DNA molecules and $C_{max}$ is its
maximum
value (equal to the inverse volume of a
DNA molecule). Below the solid curve, in segregation domain, condensed
and dissolved phases
 coexist. Thinking about very small concentrations of DNA, $C \ll C_{max}$,
we completely ignore the lower border of the segregation domain.
Above the solid curve, all DNA
molecules are in solution. A solution of long DNA molecules corresponds
to a
horizontal line $y = y_0 \ll 1$. As concentration of DNA or its length
decrease,
the window between $N_{c}$ and $N_{d}$ shrinks.
Condensation provides largest gain of free energy at $N = N_0$
where DNA molecules are neutral. This is why
the phase boundary curve peaks at $N = N_0$.
The dotted line on Fig.~2 which obeys equation $N = N_0$
devides
the plane in two
parts. At $N < N_0$ the net charge of a DNA molecule is
negative,
while
at $N > N_0$ it is positive. This net charge by definition
includes
cations which are bound to DNA with binding enery larger than $k_BT$.
Therefore, they move together
with DNA in gel electrophoresis. This means that DNA
should move in the direction opposite to the conventional one 
if $N > N_0$\ ~\cite{Shklov99}.
It would be interesting to verify this prediction experimentally. We propose
also to
combine experimental study
of the condensation phase diagram with the gel electrophoretic
measurement
of the charge on the dissolved DNA molecules.

\begin{figure}[h]
\input{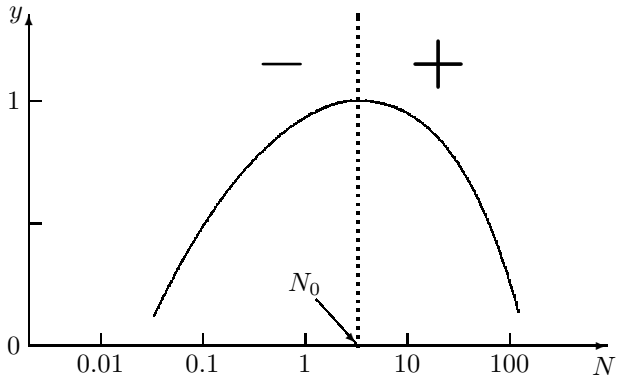}
\caption{
Phase diagram of a dilute DNA solution in the plane of
the bulk cation concentration
$N$ (in units of mM) and
variable $y$ defined by
Eq.~(\ref{y}). The solid curve was plotted using Eqs.~(\ref{boundary1})
and
(\ref{boundary2})
with fitting parameters obtained from the data~\cite{Thomas} for long
DNA screened by
Spe$^{4+}$.
 Above the curve, all DNA molecules
are in solution ($x=1$). Below the solid curve, the segregation domain
is located.
Here the condensed phase of DNA appears and
in regions far enough from the curve, it consumes most of DNA ($x \ll
1$).
The dotted line corresponds to $N=N_0$, where the net charge of DNA
changes sign from negative
to positive}
\end{figure}

\section {Criterion of equilibrium between
condensed and dissollved phases of long DNA.}

In this section we consider long DNA molecules in an aqueous solution
containing
$Z$-valent counterions of the bulk concentration $N$. DNA
condensation and resolubilization
take place when the chemical potential of DNA in its condensed and dissolved
states are equal, i.e.
\begin{equation}
\mu_{c} = \mu_{d} ~~~      .
\label{mu}
\end{equation}
For a long DNA we can neglect the entropy of DNA in both phases. We view
the
condensed phase as a bundle of
parallel molecules which stick
together
due to correlation induced
short range attraction to the nearest neigbours. The chemical
potential of DNA molecule
 in the condensed phase in this approximation is determined by %
\begin{equation}
\mu_{c} = - \varepsilon L/b ,
\label{mucond}
\end{equation}
where $L/b$ is the number of negative charges in a DNA molecule.

Each molecule of a large bundle is practically neutral. More exactly its
charge is inversly propotional to the
number of molecules in the bundle\cite{Shklov98}.
This happens because the total charge of a large bundle
keeps counterions inside the bundle very effectively.

On the other hand,
in the dissolved state, each molecule aquires finite net charge density,
$\eta^*$. This happens because a finite
 fraction of its counterions moves to the
Debye-H\"{u}ckel
atmosphere in order to increase their entropy. One can view a single DNA
molecule and its Debye-H\"{u}ckel
 atmosphere as a cylindrical capacitor with linear charge densities
$\eta^*$ (inside)
and $-\eta^*$
(outside). We recall that the net linear charge density of DNA,
$\eta^*$,
includes
the bare charge and the charge of cations residing at the very surface
of
DNA and
attached to the surface with binding energy larger than $k_BT$.

In equilibrium the electrostatic potential on the surface of DNA,
$\varphi_{0}$, is determined by the ratio of the cation
concentration in the bulk,
$N$ and the
concentration near the surface of DNA, $N_0$:
\begin{equation}
Ze\varphi_{0}= - k_BT \ln \frac{N_0}{N} \label{elpot}.
\end{equation}
This potential acts like a voltage difference applied to the capacitor.

The free energy per DNA molecule or the chemical potential of the
dissolved
phase, can be written as
\begin{equation}
\mu_{d} = - \frac{1}{2} \varphi_{0}\eta^{*} L \label{mudecond}.
\end{equation}
To derive Eq.~(\ref{mudecond}) one should add the energy of the electric
field of the capacitor, %
\begin{equation}
U = \frac{1}{2} \varphi_{0}\eta^{*} L,
\label{U}
\end{equation}
to the change of the entropy term of the free energy of cations when
they
move from the surface of DNA to the bulk of solution
\begin{equation}
{\frac{L\eta^{*}}{Ze}} k_BT~\ln \frac{N_0}{N} = -
\varphi_{0}\eta^{*} L
\label{entropy}
\end{equation}
(here Eq.~(\ref{elpot}) was used).
Equivalent derivation can be found, for example, in
Ref.~\onlinecite{Safran}.
Comparing Eq.~(\ref{mudecond}) and Eq.~(\ref{U}) we see that the change
of the
free energy of the capacitor
is equal to the energy of an electric field with a minus sign. This is
a realization of the general theorem valid for any capacitor kept at
constant voltage~\cite{Landau}.

The surface potential, $\varphi_{0}$,
can be easily related to the net charge density $\eta^*$. Indeed, at
distance $r$ from its surface, a
cylinder of radius $a$
and linear charge $\eta^*$ creates a potential
\begin{equation}
\varphi(r) = \varphi_0 - \frac{2\eta^*}{D} \ln\left(\frac{a +
r}{a}\right).
\label{phir}
\end{equation}
where $D$ is the dielectric constant of water.
This potential vanishes beyond the
Debye screening length,
\begin{equation}
r_s = (4 \pi l_B)^{-1/2} \left(N Z^{2} + ZN + 2n
\right)^{-1/2},
\label{rs}
\end{equation}
where $l_B = e^2/Dk_BT$ is the Bjerrum length. Therefore, substituting
$r= r_s$
and $\varphi(r_s) = 0$ into Eq.~(\ref{phir}) one obtains
\begin{equation}
\varphi_{0} =\frac{2\eta^*}{D} \ln \left(\frac{a+r_{s}}{a}\right).
\label{phi0}
\end{equation}
Substituting Eq.~(\ref{phi0}) into Eq.~(\ref{mudecond}) we get
\begin{equation}
\mu_{d} = - \frac{(\eta^{*})^2}{D} L~\ln (1 + r_{s}/a).
\label{mudecond1}
\end{equation}
Using the condition (\ref{mu})
together with the expressions for the chemical potentials for both
states,
Eqs.~(\ref{mucond}) and
(\ref{mudecond1}) we arrive at the final equation for the DNA reentrant
condensation
transitions
\begin{equation}
\frac{\varepsilon}{k_BT} = \frac{b(\eta^{*})^2}{D}~\ln (1 + r_{s}/a).
\label{en}
\end{equation}
To proceed further one has to know the net charge density, $\eta^{*}$.
Next section gives a review of current understanding of this quantity.

\section {Net linear charge density of screened DNA}

Conventional understanding of nonlinear screening of a strongly charged
cylinder
is based on the Poisson-Boltzmann equation.
Let us first consider a cylinder with
a negative bare linear charge density, $ - \eta$, which is
screened by $Z$-valent cations ($n= 0$). Assume that
$\eta > \eta_c$, where $\eta_c = e/l_B$ ($\eta \ = 4.2~\eta_c$ for
double
helix DNA).
Onsager and Manning~\cite{Manning} argued that such a cylinder is
partially
screened by
counterions residing at the very surface, so that the net linear charge
density of the cylinder,
 $\eta^*$, is equal to the negative universal value $-\eta_c/Z$.
The rest of the cylinder charge is linearly screened at much larger
distances
according to the linear Debye-H\"{u}ckel theory. The net charge, $\eta^*$,
does not depend
on the bulk concentration of cations, $N$ and is shown (in
units
of $-\eta_c$) by the dotted line on Fig. 3a.
The Onsager-Manning picture of condensation was confirmed by the
solution
of the Poisson-Boltzmann equation~\cite{Zimm}.

Let us now discuss a water solution, containing both
a concentration, $n$, of
monovalent salt and a concentration,
$N$  of $Z:1$, salt. When $N$
grows to some well defined
concentration, $N = N_r$,
multivalent cations replace monovalent ones at the surface of DNA.
According to conventional understanding~\cite{MN,Wilson}
this replacement of the condensed
monovalent ions changes
$- \eta^*/\eta_c$ from 1 to $1/Z$ (see the dotted line on Fig. 3b).
 In logarithmic scale this transition looks quite abrupt.

In recent papers~\cite{Perel,Shklov99} the influence of strong
correlations
of multivalent cations
at the surface of a macroion on its net charge,  $\eta^*$, was studied.
The general expression for the net linear charge density,
 $\eta^*$, was derived:
\begin{equation}
\eta^* = - \frac{\eta_c}{2Z}~\frac{\ln \left(N_0/N\right)}
{\ln (1 + r_{s}/a)}.
\label{eta*}
\end{equation}
This equation takes into account correlations with the help of the new
boundary condition
$N(r=0) = N_0$ for the Poisson-Boltzmann equation. To remind derivation
of this result let us write down the Boltzmann formula
\begin{equation}
N(r) = N_0 \exp \left( - \frac{Ze}{k_BT} (\varphi(r) - \varphi_0)
 \right) ,
\label{Nr}
\end{equation}
At $r=r_s$ the concentration $N(r)$ reaches its bulk value,
$N$.
Substituting Eq.~(\ref{phi0}) into Eq.~(\ref{Nr}) we obtain
\begin{equation}
N = N_0 \exp
 \left(\frac{2Z\eta^*}{\eta_c} \ln\left(\frac{a + r_s}{a}\right)\right).
\label{Ninf}
\end{equation}
Expressing $\eta*$ in
terms of the bulk cation concentration $N$ we arrive at
Eq.~(\ref{eta*}).
This equation was used to plot solid lines on Fig. 3.

\begin{figure}[h]
\input{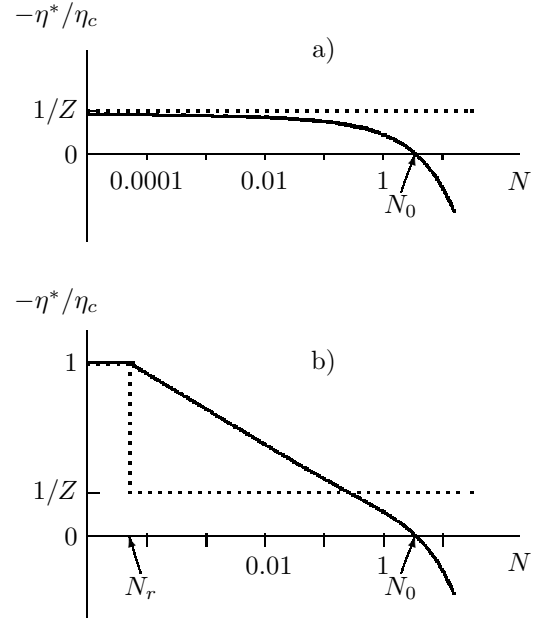}
\caption{
The dimensionless net linear charge density $\eta^*/\eta_c$ as function
of
$Z$-valent cation concentration $N$
 (in units of mM) at zero concentration $n$ of
monovalent salt (a) and at finite $n$ (b). The solid curves are
drawn
according Eq.~(\ref{eta*}) with
parameters obtained from the data~\cite{Thomas} for DNA
screened by Spe$^{4+}$. Dotted curves represent conventional
understanding of results of
Poisson-Bolzmann equation. $N_0$ is the concentration of $Z$-valent ions
at the surface of DNA and $N_r$ is the concentration $N$ at
which $Z$-valent cations replace monovalent ones.
}
\label{feta}
\end{figure}

At $n = 0$ (Fig.~3a) one can see~\cite{Shklov99} that Manning's
limiting value
of $\eta^* = - \eta_c/Z$
holds only for the unrealistically small  $N$. At larger
$N$ the absolute value of the net charge decreases,
$\eta^*$ crosses zero and becomes positive.

At finite $n$, the  screening radius, $r_s$, strongly decreases and at small
$N$ the net charge density,
 $ -\eta^*$, becomes larger than $ \eta_c/Z$
(see solid line on Fig.~3b). It was shown in Ref.~\onlinecite{Shklov99}
that when $N$ goes to $N_r$,
the density  $\eta^*$ becomes as large as $\eta_c$, so that at the
replacement point the net density, $ \eta^*$,
matches its standard value
for monovalent ions.
At larger $N$ the role of the monovalent salt is smaller and
$\eta^*$ is similar to that of $n = 0$. The density, $ \eta^*$, crosses
zero at
$N = N_0$ and becomes positive.
In the next section we will use Eq.~(\ref{eta*}) to
find the condensation and resolubilization thresholds.

\section {Reentrant condensation thresholds for long DNA molecules.
Comparison with experiment}

Now we can substitute Eq.~(\ref{eta*}) into Eq.~(\ref{en})
and obtain an explicit equation for the threshold concentrations:
\begin{equation}
\frac{\varepsilon}{k_BT} =
\frac{1}{4Z^2\xi}~\frac{\ln^2 \left(N_0/N\right)}
{\ln (1 + r_{s}/a)}.
\label{en10}
\end{equation}
Here $\xi = l_B/b$ is the conventional Manning's
parameter ($\xi = 4.2$ for DNA). Obviously, there exist two solutions,
$N_{c}$ and $N_{d}$,
for the bulk concentration, $N$,
which corresponds respectively to the
condensation and dissolution transition points.

Concentration $N_{c}$ is usually so small that $r_s$ is dominated by
monovalent ions and $r_s = r_1 =(8 \pi l_B n)^{-1/2}$.
With the help of Eq.~(\ref{en10}) we arrive at a simple equation for
$N_{c}$:
\begin{equation}
\frac{\varepsilon}{k_BT} =
\frac{1}{4Z^2\xi}~\frac{\ln^2(N_0/N_{c})} {\ln(1 + r_1/a)}.
\label{en1}
\end{equation}
To obtain an equation for $N_{d}$, we take into account that in the case of
Spe$^{4+}$ cation~\cite{Thomas},
the condensed phase dissolves at $N_{d} \approx 150$~mM.
At such a large concentration, $ r_s$ is determined by multivalent
cations.
However, if we use Eq.~(\ref{rs}), $r_s$ turns out to be much smaller
than the
average distance between cations in the bulk of solution,
which does not make physical sense.
This means that the Debye-H\"{u}ckel theory does not work.
A natural approximation in this case is to replace $ r_s$
 by average distance to closest cation
$r_d = (4\pi N_d/3)^{-1/3}$. This gives for $N_{d}$
\begin{equation}
\frac{\varepsilon}{k_BT} = \frac{1}{4Z^2\xi}~\frac{\ln^2
\left(N_0/N_{d}\right)}{\ln(1 + r_{d}/a)}.
\label{en2}
\end{equation}
In practice, the concentrations
$N_{c}$ and $N_{d}$ are known from experiments, so that important
parameters $\varepsilon$ and $N_{0}$
can be found with the help of Eqs.~(\ref{en1}) and (\ref{en2}).
For example, $N_0$ can be found by eliminating $\varepsilon$
from eqs.(\ref{en1}, \ref{en2}).
In the specific experimental situation of Ref.~\onlinecite{Thomas}
where long DNA  molecules condensed with a 4-valent cation,
Spe$^{4+}$, $N_c = 0.025$~mM and $N_d = 150$~mM at
the lowest concentration $n = 10$~mM of NaCl.
Using these values we obtain from Eqs.~(\ref{en1}), (\ref{en2})
$N_0 = 3.2$~mM and $\varepsilon/k_BT \approx 0.07$.
The last value favorably agrees with the energy of attraction between
the
CoHex condensed DNA obtained in the osmotic stress experiment
\cite{Rau92} $\varepsilon = 0.08~k_BT$.

Knowing $N_0$ and $\varepsilon$, we can try to
reproduce the experimental dependence of the condensation threshold $N_c$
on concentration of NaCl, obtained in Ref.~\onlinecite{Thomas}.
 This can be done with the help of Eq.~(\ref{en1}):
\begin{equation}
\ln N_c = \ln N_0  - \left(4Z^2\xi~ \frac{\varepsilon}{k_BT} ~
\ln(1+ r_1/a)\right)^{1/2}.
\label{Nc}
\end{equation}
The calculated function $N_c(n)$ is shown on Fig. 4
together with
the experimental points from Ref.~\onlinecite{Thomas}. It is clear that
Eq.~(\ref{Nc}) closely reproduces experimental behavior till
 $n \lesssim 50$~mM. The later value is close to the
concentration of monovalent cations which
is needed to replace $4$-valent cations at the surface of DNA
for $N \approx 0.01$~mM.
It seems that at $n > 50$ mM, condensation happens simultaneously with
the replacement of
monovalent cations by multivalent ones, $i.e.$ at $N_c \sim N_r$, as was
assumed in Ref.~\onlinecite{Wilson}.

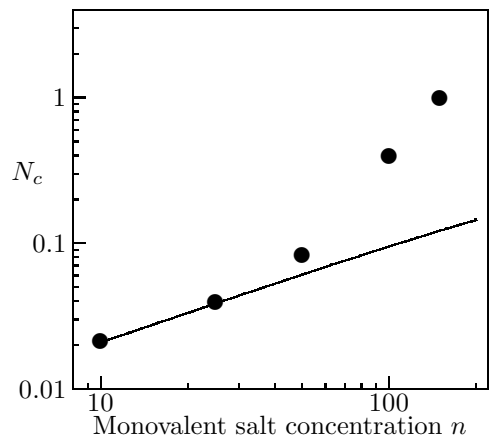
\begin{figure}[h]
\setlength{\unitlength}{0.120900pt}
\ifx\plotpoint\undefined\newsavebox{\plotpoint}\fi
\sbox{\plotpoint}{\rule[-0.200pt]{0.400pt}{0.400pt}}%
\begin{picture}(1500,1450)(-100,-100)
\font\gnuplot=cmr10 at 10pt
\gnuplot
\sbox{\plotpoint}{\rule[-0.200pt]{0.400pt}{0.400pt}}%
\put(120,123){\makebox(0,0)[r]{0.01}}
\put(140.0,123.0){\rule[-0.200pt]{4.818pt}{0.400pt}}
\put(140.0,260.0){\rule[-0.200pt]{2.409pt}{0.400pt}}
\put(140.0,341.0){\rule[-0.200pt]{2.409pt}{0.400pt}}
\put(140.0,398.0){\rule[-0.200pt]{2.409pt}{0.400pt}}
\put(140.0,442.0){\rule[-0.200pt]{2.409pt}{0.400pt}}
\put(140.0,478.0){\rule[-0.200pt]{2.409pt}{0.400pt}}
\put(140.0,509.0){\rule[-0.200pt]{2.409pt}{0.400pt}}
\put(140.0,535.0){\rule[-0.200pt]{2.409pt}{0.400pt}}
\put(140.0,558.0){\rule[-0.200pt]{2.409pt}{0.400pt}}
\put(120,579){\makebox(0,0)[r]{0.1}}
\put(140.0,579.0){\rule[-0.200pt]{4.818pt}{0.400pt}}
\put(140.0,716.0){\rule[-0.200pt]{2.409pt}{0.400pt}}
\put(140.0,797.0){\rule[-0.200pt]{2.409pt}{0.400pt}}
\put(140.0,854.0){\rule[-0.200pt]{2.409pt}{0.400pt}}
\put(140.0,898.0){\rule[-0.200pt]{2.409pt}{0.400pt}}
\put(140.0,934.0){\rule[-0.200pt]{2.409pt}{0.400pt}}
\put(140.0,965.0){\rule[-0.200pt]{2.409pt}{0.400pt}}
\put(140.0,991.0){\rule[-0.200pt]{2.409pt}{0.400pt}}
\put(140.0,1014.0){\rule[-0.200pt]{2.409pt}{0.400pt}}
\put(120,1035){\makebox(0,0)[r]{1}}
\put(140.0,1035.0){\rule[-0.200pt]{4.818pt}{0.400pt}}
\put(140.0,1173.0){\rule[-0.200pt]{2.409pt}{0.400pt}}
\put(140.0,1253.0){\rule[-0.200pt]{2.409pt}{0.400pt}}
\put(140.0,1310.0){\rule[-0.200pt]{2.409pt}{0.400pt}}
\put(140.0,123.0){\rule[-0.200pt]{0.400pt}{2.409pt}}
\put(186.0,123.0){\rule[-0.200pt]{0.400pt}{2.409pt}}
\put(227,82){\makebox(0,0){10}}
\put(227.0,123.0){\rule[-0.200pt]{0.400pt}{4.818pt}}
\put(499.0,123.0){\rule[-0.200pt]{0.400pt}{2.409pt}}
\put(658.0,123.0){\rule[-0.200pt]{0.400pt}{2.409pt}}
\put(771.0,123.0){\rule[-0.200pt]{0.400pt}{2.409pt}}
\put(858.0,123.0){\rule[-0.200pt]{0.400pt}{2.409pt}}
\put(930.0,123.0){\rule[-0.200pt]{0.400pt}{2.409pt}}
\put(990.0,123.0){\rule[-0.200pt]{0.400pt}{2.409pt}}
\put(1043.0,123.0){\rule[-0.200pt]{0.400pt}{2.409pt}}
\put(1089.0,123.0){\rule[-0.200pt]{0.400pt}{2.409pt}}
\put(1130,82){\makebox(0,0){100}}
\put(1130.0,123.0){\rule[-0.200pt]{0.400pt}{4.818pt}}
\put(1402.0,123.0){\rule[-0.200pt]{0.400pt}{2.409pt}}
\put(140.0,123.0){\rule[-0.200pt]{156.929pt}{0.400pt}}
\put(1439.0,123.0){\rule[-0.200pt]{0.400pt}{144pt}}
\put(140.0,1310.0){\rule[-0.200pt]{156.929pt}{0.400pt}}
\put(789,10){\makebox(0,0){Monovalent salt concentration $n$}}
\put(0,800){\makebox(0,0){$N_c$}}
\put(140.0,123.0){\rule[-0.200pt]{0.400pt}{144pt}}
\put(227,270){\usebox{\plotpoint}}
\multiput(227.00,270.58)(1.528,0.497){57}{\rule{1.313pt}{0.120pt}}
\multiput(227.00,269.17)(88.274,30.000){2}{\rule{0.657pt}{0.400pt}}
\multiput(318.00,300.58)(1.461,0.497){59}{\rule{1.261pt}{0.120pt}}
\multiput(318.00,299.17)(87.382,31.000){2}{\rule{0.631pt}{0.400pt}}
\multiput(408.00,331.58)(1.511,0.497){57}{\rule{1.300pt}{0.120pt}}
\multiput(408.00,330.17)(87.302,30.000){2}{\rule{0.650pt}{0.400pt}}
\multiput(498.00,361.58)(1.461,0.497){59}{\rule{1.261pt}{0.120pt}}
\multiput(498.00,360.17)(87.382,31.000){2}{\rule{0.631pt}{0.400pt}}
\multiput(588.00,392.58)(1.528,0.497){57}{\rule{1.313pt}{0.120pt}}
\multiput(588.00,391.17)(88.274,30.000){2}{\rule{0.657pt}{0.400pt}}
\multiput(679.00,422.58)(1.511,0.497){57}{\rule{1.300pt}{0.120pt}}
\multiput(679.00,421.17)(87.302,30.000){2}{\rule{0.650pt}{0.400pt}}
\multiput(769.00,452.58)(1.511,0.497){57}{\rule{1.300pt}{0.120pt}}
\multiput(769.00,451.17)(87.302,30.000){2}{\rule{0.650pt}{0.400pt}}
\multiput(859.00,482.58)(1.511,0.497){57}{\rule{1.300pt}{0.120pt}}
\multiput(859.00,481.17)(87.302,30.000){2}{\rule{0.650pt}{0.400pt}}
\multiput(949.00,512.58)(1.581,0.497){55}{\rule{1.355pt}{0.120pt}}
\multiput(949.00,511.17)(88.187,29.000){2}{\rule{0.678pt}{0.400pt}}
\multiput(1040.00,541.58)(1.564,0.497){55}{\rule{1.341pt}{0.120pt}}
\multiput(1040.00,540.17)(87.216,29.000){2}{\rule{0.671pt}{0.400pt}}
\multiput(1130.00,570.58)(1.620,0.497){53}{\rule{1.386pt}{0.120pt}}
\multiput(1130.00,569.17)(87.124,28.000){2}{\rule{0.693pt}{0.400pt}}
\multiput(1220.00,598.58)(1.620,0.497){53}{\rule{1.386pt}{0.120pt}}
\multiput(1220.00,597.17)(87.124,28.000){2}{\rule{0.693pt}{0.400pt}}
\multiput(1310.00,626.58)(1.700,0.497){51}{\rule{1.448pt}{0.120pt}}
\multiput(1310.00,625.17)(87.994,27.000){2}{\rule{0.724pt}{0.400pt}}
\sbox{\plotpoint}{\rule[-0.500pt]{1.000pt}{1.000pt}}%
\put(227,274){\circle*{50}}
\put(587,397){\circle*{50}}
\put(858,543){\circle*{50}}
\put(1130,853){\circle*{50}}
\put(1289,1035){\circle*{50}}
\put(227,274){\usebox{\plotpoint}}
\end{picture}
\caption {
The condensation threshold $N_c$ as a function of monovalent salt
concentration, $n$.
The solid curve corresponds to Eq.~(\ref{Nc}).
Experimental data~\cite{Thomas} are shown by the large dots.}
\end{figure}

We would like to emphasize that in the broad range
of lower monovalent salt concentrations $n < 50$~mM
when multivalent cations cover the DNA surface, condensation does not
happen.
In this range, the Coulomb repulsion of negative DNA molecules
 is strong enough to prevent condensation because,
as shown on Fig.~3b, the absolute value of the net negative
charge on the polymer,
$- \eta^*$, does not abrubtly decrease from $\eta_c$ to $\eta_c/Z$ at
$N_c
\sim N_r$, but still stays larger than
$\eta_c/Z$ in the large interval of $N > N_r$.

In contrast with $N_c$, the resolubilization threshold $N_d$ practically
does not
depend on $n$. This happens because at $
N = N_d$, screening is dominated by multivalent
cations and agrees with our Eq.~(\ref{en2}).

Let us now discuss another experimental study\cite{Livolan96} in which
DNA
were condensed with trivalent Spd$^{3+}$ cations.
In this case, for $n = 4$~mM thresholds are equal
 $N_c \approx 2$~mM and $N_d \approx 60$~mM, which yields
$N_0 \approx 11$~mM and $\varepsilon / k_BT \approx 0.02$.
This value of $\varepsilon / k_BT $  is about three times smaller than for
Spe$^{4+}$. This is quite reasonable, if attraction has an electrostatic
nature.

The concentrations, $N_0$, we obtained above,
are substantially larger than the microscopic theory~\cite{Shklov99}
predicts
for the point-like 3- and 4-valent cations.
 Actually, cations Spd$^{3+}$ and Spe$^{4+}$
are quite long ($\approx 15$ and $20~\AA$ respectively) linear polymers.
Their length is approximately equal to the
 average distance between cation centers
in two-dimensional strongly correlated liquid of cations on the surface
of DNA.
Thus, to obtain a more reliable theoretical prediction of $N_0$, one has to
study the thermodynamic properties of the strongly
 correlated (possibly nematic) liquid of these ions on an uniform
negative background by numerical methods. We will address this problem
in
our next paper.

The application of our theory to experimental data~\cite{Thomas,Livolan96}
is not
completely convincing, because of the following. Our theory assumes that the
concentration of DNA is so small
 that the concentration of DNA phosphates
$N_{ph}$ is smaller than $N$, so that the DNA charge can always be
compensated by Z-valent cations.
 Actually in experimental conditions~\cite{Thomas,Livolan96} the
condensation threshold $N_c$ happens to
be close to $N_{ph}$. It is possible, therefore, that $N_c$ in this case
is
determined by the condition
$N_c \sim N_{ph}$ rather than the repulsion of DNA helices. This
means
that the actual value of $N_c$
can be lower than the measured one. Therefore, the true values of $N_0$
should be
somewhat smaller, and that of
$\varepsilon$ should be somewhat larger than our estimates. We suggest
a repetition of experiments at smaller
 concentrations of long DNA.

Another way to improve the reliability of the extraction of the value of $N_0$
from
experiment is to study the condensation
 of short DNA (see next section).

Note that Ref.~~\onlinecite{Thomas} also contains information on the
single and triple
stranded DNA helices, which condensed in the narrower and wider range of
[Spe$^{4+}$]
than the double helices respectively. This tendency agrees with idea that
correlations
play the major role in this phenomenon.

\section {Condensation of short DNA molecules.}

In this Section we explicitly deal with the mixing entropy of
DNA molecules in the dissolved state, and calculate the
phase diagram for the DNA solution.
We consider double helix DNA molecules of length $L$,
with $L/b$ bases each.

The free energy per DNA molecule in solution with
concentration $C$ is
\begin{equation}
F = - x k_BT \left( \ln \left(\frac{C_{max}}{Cx}\right) + 1 \right) -
(1-x) \frac{L}{b}\Delta,
\label{F}
\end{equation}
where $1 - x$ is the fraction of DNA molecules in the condensate. The
first term in
Eq.~(\ref{F}) is the entropy of the dissolved DNA phase per molecule of
DNA. Here
$C_{max} \sim 1/\pi a^2 L$ is the inverse volume of the DNA molecule.
The second term in Eq.~(\ref{F}) is the average free
energy per molecule in the condensed state.
Here $\Delta$ is the difference
between the energy of short range attraction $\varepsilon$
and the free energy $b\mu_d/L$ of the screening
atmosphere per nucleotide (see Eq.~(\ref{mudecond1})):
\begin{equation}
\frac{\Delta}{k_BT} = \frac{\varepsilon}{k_BT} -
\frac{1}{4Z^2 \xi} \frac{\ln^2 ( N_0 / N )}{\ln (1+r_s/a)}.
\label{Den}
\end{equation}
Minimizing Eq.~(\ref{F}) with respect of $x$, we find
\begin{equation}
x = \frac{C_{max}}{C}\exp\left(- \frac{\Delta}{k_BT}~\frac{L}{b}\right).
\label{x}
\end{equation}
Then the boundary of two-phase domain $x=1$ corresponds to the
condition
\begin{equation}
\ln \frac{C_{max}}{C} = \frac{\Delta}{k_BT}~\frac{L}{b},
\label{boundary}
\end{equation}
which can be explicitly written for $N < N_0$ as
\begin{equation}
\frac{b k_BT}{L \varepsilon} \ln {\frac{C_{max}}{C}} =
1 - \frac{k_BT}{\varepsilon}\frac{1}{4Z^2 \xi} \frac
{ \ln^{2}(N_0/N)}{\ln(1 + r_1/a)},
\label{boundary1}
\end{equation}
and for $N > N_0$
\begin{equation}
\frac{bk_BT}{L \varepsilon} \ln \frac{C_{max} }{C}=
1 - \frac{k_BT}{\varepsilon}
 \frac{1}{4Z^2 \xi} \frac { \ln^{2}(N_0/N)}{\ln(1 + r_N/a)}.
\label{boundary2}
\end{equation}
where $r_N = (4\pi N/3)^{-1/3}$. The last three equations specify the
conditions when the chemical potentials of the condensed and dissolved states of DNA
are equal. The curve described by Eqs.~(\ref{boundary1}) and (\ref{boundary2}),
forms a boundary on the phase diagram, separating the region of the
dissolved DNA (above it) from the region where the condensed and
dissolved  phases of DNA coexist (below it). Within
the separation domain the fraction of dissolved DNA is given by
Eq.~(\ref{x}).

The position and the shape of the phase boundary
 obviously depends on the values of the parameters
$N_0$, $\varepsilon / k_BT $ and $2Z^2 \xi$. For double-helix DNA
with Spe$^{4+}$ we find $2Z^2 \xi = 134$ and using the estimates
$\varepsilon / k_BT=0.07$ and $N_0 \approx 3.2$~mM obtained
above we can calculate the phase boundary
for condensation of the DNA fragments of the arbitrary length. This
boundary is shown on Fig.~2 by the solid line.

The dotted line $N = N_0$
divides the phase diagram of Fig. 2 into two regions where the DNA
molecules have opposite signs. Positive net charge of dissolved DNA
can be measured in an electrophoresis experiment because cations
included
in this charge are bound to DNA with a binding energy larger than $k_BT$
and therefore
move together with DNA molecule. At $N > N_0$ one should
see that DNA molecules are positive
both above and below phase boundary of Fig.~2.
However, below the boundary, intensity of corresponding
electrophoresis peak should decay rapidly
with the distance from the boundary.
This intensity is picked up by slowly moving bundles
of DNA molecules, which at large enough $N$ can be also
positive.

\section {Conclusion}

The theory of DNA condensation and resolubilization by multivalent
cations
presented above makes several novel, well-defined predictions which have
not been confirmed by experiment yet. The main results of the study are
summarized
in the phase diagram presented on Fig. 2. There are only two physical
parameters $\varepsilon$ and $ N_0$ on which the shape of the phase
boundary depends. Therefore measuring just a few threshold
concentrations of
multivalent cations $N_c$ and $N_d$ for solution of DNA fragments of
different length
and/or concentrations should yield several independent determinations of
these
quantities, and at the same time provide the test for the
selfconsistency
of our model. Experimental studies of DNA phase diargam with different
multivalent cations would provide the values of the attractive energy
for different
ions. In the present theory the origin of parameters $\varepsilon$ and $
N_0$
was not specified and they were treated as phenomenological parameters.
Comparison of experimentally determined
$\varepsilon$ and $ N_0$ values for different counterions  should yied
information about the nature of the attraction. It is worth noting here
that  if the attraction is of electrostatic
correlation origin the value of $\varepsilon$ in principle
includes attraction due to the correlations of multivalent cations as
well as
all of the repulsive non-Coulomb DNA-DNA interaction. However, the first
interaction decays slower than the second one. Then the binding energy
$\varepsilon$ is determined by correlations. In this case,
the two quantities $\varepsilon$ and $N_0$ are not
independent~\cite{Shklov98,Shklov99}
namely $\ln{N_0} \propto \varepsilon$. For the model of the uniformly
charged cylinder
both parameters were calculated
microscopically~\cite{Shklov98,Shklov99}.

We would like to emphasize that the concentration $N_0$ found in this
paper plays
extremely important role in any phenomenon related to screening
of DNA molecules by multivalent ions~\cite{Perel,Shklov99}. In this
paper, we
try to attract attention to the fact that $N_0$ plays major role in
electrophoresis,
because the net linear charge density of DNA
$\eta^*$ changes sign at $N = N_0$.
It was predicted~\cite{Perel,Shklov99} that DNA should start
moving in the opposite direction at $N > N_0$.
It is not obvious that one can see
this phenomenon for long DNA. Indeed, in large interval of
concentrations $ N_0 < N < N_d$
most of long DNA molecules are condensed in low mobilty bundles while
concentrations
$N > N_d$ may be difficult for experiment because of large
dissipation of heat. Therefore, we suggest doing electrophoresis 
of a solution of short DNA fragments. In this case all DNA molecules
have unconventional sign of mobility at smaller than $N_d$
concentrations.
The phase diagram shown in Fig. 2 predicts good conditions for such an
experiment.
It would be very interesting to verify predicted correlations between
the 
reentrant condensation and unconventional electrophoresis.

\acknowledgements

We are grateful to A. Yu. Grosberg, V. A. Bloomfield and R. Podgornik
for valuable discussions. This work was supported by NSF DMR-9616880 (T. N. and B.S)
and NIH GM 28093 (I. R.).

\end{multicols}
\end{document}